\documentclass[preprint,12pt,authoryear]{elsarticle}

\usepackage{amssymb}

\usepackage{lineno}
\usepackage{amsmath, tabu}
\usepackage[usenames, dvipsnames]{color}
\usepackage{amstext} 
\usepackage{array}   
\usepackage{graphicx}
\usepackage{caption}
\usepackage{subcaption}

\journal{Journal}

\begin{document}

\begin{frontmatter}



\title{A Multi-Stencil Fast Marching Method with Path Correction for Efficient Reservoir Simulation and Automated History Matching}


\address[1]{State Key Laboratory of Oil and Gas Reservoir Geology and Exploitation, Southwest Petroleum University, Chengdu, Sichuan Province, 610500, China}
\address[2]{Research Centre for Mathematics and Interdisciplinary Sciences, Shandong University, Qingdao, Shandong Province, 266237, China}
\address[3]{School of Petroleum Engineering, China University of Petroleum, Qingdao, Shandong Province, 266580, China}
\address[4]{School of Science, Qingdao University of Technology, Qingdao, Shandong Province, 266033, China}
\author[1,2]{Zhao Zhang\corref{cor1}}
\cortext[cor1]{Corresponding author}
\ead{zhaozhang@sdu.edu.cn}
\author[1]{Jianchun Guo\corref{cor1}}
\ead{guojianchun@vip.163.com}
\author[3]{Kai Zhang}
\author[4]{Piyang Liu}
\author[3]{Xia Yan}

\begin{abstract}
The efficiency of reservoir simulation is important for automated history matching (AHM) and production optimization, etc. The fast marching marching method (FMM) has been used for efficient reservoir simulation. FMM can be regarded as a generalised streamline method but without the need to construct streamlines. In FMM reservoir simulation, the Eikonal equation for the diffusive time-of-flight (DTOF) is solved by FMM and then the governing equations are computed on the 1D DTOF coordinate. Standard FMM solves the Eikonal equation using a 4-stencil algorithm on 2D Cartesian grids, ignoring the diagonal neighbouring cells. In the current study, we build a 8-stencil algorithm considering all neighbouring cells, and use local analytical propagation speeds. In addition, a local path-correction coefficient is introduced to further increase the accuracy of DTOF solution. Next, a discretisation scheme is built on the 1D DTOF coordinate for efficient reservoir simulation. The algorithm is validated on homogeneous and heterogeneous test cases, and its potential for efficient forward simulation in AHM is demonstrated by two examples of dimensions 2 and 6. 

\end{abstract}


\end{frontmatter}


\section{Introduction}
\label{intro}
Hydrocarbon reservoir models are usually heterogeneous but only sparse hard data from core samples and well logging are available for building reservoir models. Although there is soft data from seismic inversion, the accuracy is generally low.  In consequence, geological models have inherent uncertainty, and the evaluation of uncertainty is crucial for production prediction as well as development optimization \citep{oliver2008inverse}. For scenarios where well testing or production data is available, automated history matching (AHM) can be adopted to reduce the uncertainty of geological models, and reservoir simulation typically needs to be conducted on a large number of realisations to explore how dynamic responses are affected by the change of model parameters \citep{oliver2011recent, arnold2019uncertainty, demyanov2019uncertainty, zhang2021history}.
However, conventional reservoir simulation methods based on finite difference, finite volume or finite element methods \citep{zhang2018workflow, zhang2021numerical} are generally time-consuming for transient problems, and it can be prohibitively expensive to run simulations on all realisations in the process of AHM. Proxies can be used in place of reservoir simulations to obtain dynamic responses \citep{zhao2020classification}, but proxies by interpolation are based on reservoir simulation results while those by reduced physics \citep{zhang2017tracing} are low in accuracy for complex flow mechanisms. Therefore, efficient reservoir simulation methods are of great importance.

The streamline method is an efficient reservoir simulation approach \citep{datta2007streamline}. Its basic idea is to build streamlines based on pressure field obtained by finite difference methods, and then simulate saturation efficiently on streamlines along the time-of-flight (TOF) coordinate efficiently. Streamlines need to be updated once the pressure field is changed. Since the pressure field varies much less than the saturation field, the time step for pressure can be much larger than that for saturation. Generation of streamlines is often time-consuming and even difficult for complex geometries and unstructured grids \citep{matringe2008, hagland2009, rasmussen2010, klausen2012, TEIXEIRA2021108369}, while the mapping of saturation field between streamlines and grid cells causes extra inaccuracies. Given these advantages and disadvantages, streamline methods have been applied in many practical problems \citep{thiele2003, batycky2005, ZHANG2021108617}.

The fast marching method (FMM) is another efficient reservoir simulation approach \citep{sharifi2014}. The basic idea is to write the pressure diffusion equation in the frequency domain by Fourier transform and obtain the Eikonal equation of the diffusive time-of-flight (DTOF) for the pressure wave in the high-frequency limit \citep{vasco2000}. Then the Eikonal equation is solved by FMM for DTOF. Next, DTOF is used as a spatial coordinate and the governing equations can be mapped onto the 1D DTOF coordinate for discretisation and simulation \citep{zhang2016}. FMM has been applied for numerical well testing, geological model ranking, reservoir simulation and automated history matching \citep{xie2015,YOUSEFZADEH2021108620, TENG2020107183}.   
FMM can be regarded as a generalisation of the streamline method as the DTOF contours are perpendicular to streamlines, and solving the governing equations on the DTOF coordinate is akin to solving along a group of streamlines. Compared  to streamline methods, the benefit of FMM is that there is no need to trace and build streamlines.

The standard FMM by \cite{sethian1996} for solving the Eikonal equation is of 4-stencil in 2D structured grids. However, there are 8 neighbouring cells for each internal cell in a Cartesian grid. The ignorance of diagonal neighbours in standard FMM leads to inaccuracies. There have been studies on improving the accuracy of FMM. The higher-accuracy FMM \citep{sethian1999level} improves the accuracy of FMM by approximating the gradient by a second-order scheme, but the scheme is not monotone and a check-and-correct step is included in their algorithm. \citet{hassouna2007multistencils} developed a multistencil FMM by using directional derivatives for diagonal cells. \citet{yoon2017} combined the multistencil FMM with the second-order scheme for higher accuracy. These studies concerns about FMM on Cartesian grids. For FMM on unstructured grids, please refer to \citet{zhang2020fast} for a discussion.    
 
In the current study, we will develop a multistencil FMM scheme that uses local analytical propagation speed for solving the Eikonal equation and reservoir simulation on Cartesian grids which is widely adopted in geological modelling. The path correction method \citep{zhang2020fast} is employed to compensate the difference between the FMM path on the grid and the characteristic curve. We will validate the new algorithm on homogeneous and highly heterogeneous reservoirs and demonstrate its potential for automated history matching. The CPU time needed can be several orders of magnitude less than conventional finite difference or finite element method.

This paper is organised as follows. First, the background of using FMM in reservoir simulation is reviewed. Second, a new multistencil FMM with path correction for the Eikonal equation of DTOF is proposed and a discretisation scheme on the 1D DTOF coordinate is developed. Third, the algorithm is validated on homogeneous and heterogeneous examples. Fourth, the efficiency of the algorithm for AHM is demonstrated. 

\section{Background of FMM for Reservoir Simulation}

The governing equation for single-phase slightly compressible flow in a petroleum reservoir is governed by

\begin{equation}
\phi\mu c_t\frac{\partial p}{\partial t}=\nabla \cdot(K\nabla p)+q~, \label{pt}
\end{equation}
where $p$ is pressure, $K$ is permeability, $c_t$ is total compressibility, $\mu$ is viscosity, $\phi$ is porosity and  $q$ is the source/sink term ($s^{-1}$). By Fourier transform, the asymptotic solution for Eq.~\eqref{pt} can be derived to show that the DTOF for pressure front propagation in the high frequency limit is governed by the Eikonal equation

\begin{equation}
|\nabla \tau|=1/f~,\label{eikonal2}
\end{equation}
where $\tau$ is DTOF and the propagation speed $f$ is computed as

\begin{equation}
f=\sqrt{\frac{K}{\phi\mu c_t}}~.
\end{equation}
Then FMM is used to solve  Eq.~\eqref{eikonal2} for $\tau$. Next, $\tau$ can be used as a spatial coordinate, and the governing equation \eqref{pt} can be mapped onto the 1D $\tau$-coordinate to be solved efficiently. We have

\begin{equation}
\omega(\tau)\frac{\partial P}{\partial t}=\frac{\partial}{\partial \tau}\left[\omega(\tau)\frac{\partial P}{\partial \tau}\right]+q ~, \label{taueq}
\end{equation}
with flow-rate boundary condition 

where $\omega(\tau)$ is computed as
\begin{equation}
	\omega(\tau)=\frac{\partial V(\tau)}{\partial \tau}~, \label{omega}
\end{equation}
where $V(\tau)$ is the drainage volume by DTOF. The entire algorithm consisting of solving the Eikonal equation for DTOF and solving the flow equation on the 1D DTOF coordinate can be named \textbf{FMM-DTOF}.

\section{A Multistencil FMM with Path Correction for Reservoir Simulation}
\begin{figure}[!htp] 
	\centering
	\begin{subfigure}[t]{0.3\textwidth}
		\centering
		\includegraphics[width=1\textwidth]{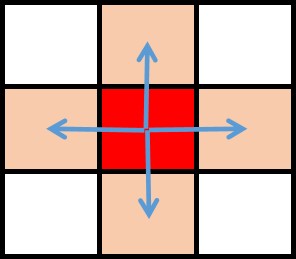}
		\caption{}
		\label{four} 
	\end{subfigure}
	\begin{subfigure}[t]{0.3\textwidth}
		\centering
		\includegraphics[width=1\textwidth]{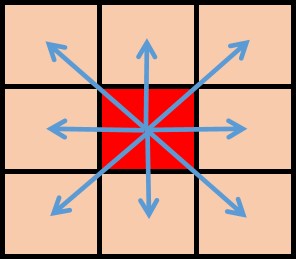}
		\caption{}
		\label{eight}
	\end{subfigure}
	\caption{Standard FMM is of 4-stencil ignoring diagonal neighbouring cells (a); our algorithm for solving th Eikonal equation considers all neighbouring cells and is of 8-stencil (b).}
\end{figure}

\begin{figure}[!htp]
	\centering
	\includegraphics[width=0.3\linewidth]{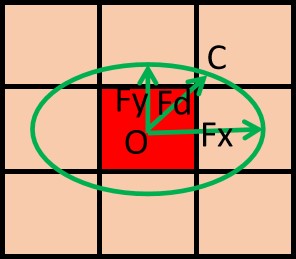}
	\caption{The local propagation speed profile is an ellipse with two axes being $F_x$ and $F_y$. The magnitude of the local propagation speed to any direction can be calculated analytically as the length of the directional vector from the cell centre O to the intersection point C.}
	\label{fd}
\end{figure}

The standard FMM by \cite{sethian1996} for solving Eq.~\eqref{eikonal2} only considers the propagation from a cell to its four neighbouring cells (Fig.~\ref{four}). Motivated by \citet{hassouna2007multistencils}, we consider all of its eight neighbours in a 2D Cartesian grid (Fig.~\ref{eight}). Motivated by \citet{zhang2020fast}, local analytical propagation speeds are calculated as in Fig.~\ref{fd}. For a cell centred at the red O, the local propagation speed profile is the ellipse with two axes being $F_x$ and $F_y$, which are the speed in $+x$ and $+y$ directions. the propagation speed $f_d$ from the cell to its upper-right neighbour is the length of the vector from O to the intersection point C. Analytical propagation speed $f_d$ to other neighbours can be calculated similarly. Then Eq.~\eqref{eikonal2} is solved as

\begin{equation}
	\Delta \tau=l/f_d~,\label{dis0}
\end{equation}
where $l$ is the distance between the centres of two neighbouring cells, and $\Delta \tau$ is the increment of $\tau$ between two neighbouring cells.
Apart from considering eight neighbours and using local analytical propagation speeds, other details in our algorithm is the same as the standard FMM by \cite{sethian1996} whose core is the establishment of a narrow band and choose the cell with lowest $\tau$ as the propagation direction. The multistencil FMM method (MSFMM) is therefore
\begin{enumerate}
	\item Label all boundary cells as frozen. These cells have $\tau=0$.
	\item Compute $\tau$ for all cells that have at least one frozen neighbour and label them as candidate. All eight neighbours are considered. All candidate cells form the narrow band.
	\item Find the cell with smallest $\tau$ in the narrow band, mark it frozen and remove it from narrow band. 
	\item Solve $\tau$ for all neighbours of the recently frozen cell and move them to the narrow band. If a neighbour is already in the narrow band, it is recomputed.
	\item Return to step 3. The loop continues until all cells become frozen. 
\end{enumerate}

\begin{figure}[!htp]
	\centering
	\includegraphics[width=0.4\linewidth]{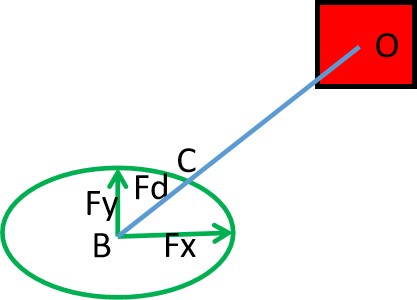}
	\caption{For homogeneous cases, the global speed profile is an ellipse centred at B with axes $F_x$ and $F_y$. B denotes the boundary for FMM, e.g. the wellbore. The propagation speed $F_d$ is the length of BC where C is the intersection point of BO and the speed ellipse. Then the analytical propagation time from B to O is simply the length of BO divided by $F_d$.}
	\label{ana}
\end{figure}

MSFMM using Eq.~\eqref{dis0} and local analytical propagation speed is of first-order accuracy and only one upstream cell is involved for computing a downstream cell. The benefit is that the scheme is guaranteed monotone so no causality check-and-correct step is needed. In addition, since local analytical propagation speed is used, the algorithm should be vary accurate if the propagation path is correct. However, the propagation path on a grid is usually different from the characteristic curve, i.e. the correct path, which is the main cause of inaccuracy.  For a homogeneous model, the propagation time can be solved exactly by an analytical method. As in Fig.~\ref{ana}, for an arbitrary cell centred at O, a line BO connects the boundary B which is the starting cell of propagation and O. The speed profile is an ellipse centred at B with axes $F_x$ and $F_y$. The propagation speed $F_d$ is the length of BC where C is the intersection point of BO and the speed ellipse. Then the analytical propagation time is simply $\tau_{ana}=\frac{BO}{F_d}$. On the other hand, the corresponding propagation time $\tau_{num}$ can be solved numerically by MSFMM. Then a correction factor is defined as 

\begin{equation}
	C_i=\frac{\tau_{ana}}{\tau_{num}}
\end{equation}
where $C_i$ is local for each cell and $i$ is the cell number on the grid. It is obvious that $C_i*\tau_{num}=\tau_{ana}$ for homogeneous cases. The purpose of $C_i$ is to compensate the difference between the numerical path on a grid and the characteristic curve. For heterogeneous cases, calculating the actual path along the characteristic curve is difficult, and we use $C_i$ obtained on a homogeneous model directly as an approximation. This is approximately correct if the grid resolution is close to uniform. 

The steps of our multistencil fast marching method with path correction (MSFMMC) for solving Eq.~\eqref{eikonal2} is as follows.
\begin{enumerate}
	\item If the model is homogeneous, compute the analytical solution directly.
	\item If the model is heterogeneous, do steps 3-5.
	\item Assume a homogeneous model, compute both the analytical and numerical propagation time by MSFMM and then calculate the correction factor $C_i$ for each grid cell.
	\item Compute the numerical propagation time $\tau_{num}$ by MSFMM for the heterogeneous model.
	\item For each cell $i$, the propagation time is obtained as $\tau=C_i*\tau_{num}$.
\end{enumerate}

Next, the finite difference/volume method is used to solve Eq.~\eqref{taueq} on 1D DTOF coordinate. Our discretisation scheme is presented here. First, Eq.~\ref{omega} is implemented into Eq.~\ref{taueq} to have

\begin{equation}
	\frac{\partial V(\tau)}{\partial \tau}\frac{\partial P}{\partial t}=\frac{\partial}{\partial \tau}\left[\frac{\partial V(\tau)}{\partial \tau}\frac{\partial P}{\partial \tau}\right]+\frac{q}{c_t} ~, \label{alleq}
\end{equation}
which is discretised on the 1D DTOF coordinate. The DTOF coordinate axis is split into uniform grid cells, i.e. intervals. For an internal cell on the 1D grid, the discretised equation is

\begin{equation}
	\frac{V_{i+1}-V_{i-1}}{2d\tau}\frac{P_i^{n+1}-P_i^n}{dt}d\tau=\frac{V_{i+1}-V_{i}}{d\tau}\frac{P_{i+1}^{n}-P_i^n}{d\tau}-\frac{V_{i}-V_{i-1}}{d\tau}\frac{P_{i}^{n}-P_{i-1}^n}{d\tau}~, \label{dis1}
\end{equation}
where $i$ is the cell number and $n$ is the time step number; for the first cell on the 1D grid, we have

\begin{equation}
 	\frac{V_{1}-V_{0}}{d\tau}\frac{P_0^{n+1}-P_0^n}{dt}d\tau
 	=\frac{V_{1}-V_{0}}{d\tau}\frac{P_{1}^{n}-P_0^n}{d\tau}+\frac{qV_{0}}{c_t}~,\label{dis2}
\end{equation}
where $0$ denotes the first cell; for the last cell on the 1D grid, we have

\begin{equation}
	\frac{V_{m}-V_{m-1}}{d\tau}\frac{P_m^{n+1}-P_m^n}{dt}d\tau
	=-\frac{V_{m}-V_{m-1}}{d\tau}\frac{P_{m}^{n}-P_{m-1}^n}{d\tau}~.\label{dis3}
\end{equation}
where $m$ is the total number of grid cells. The algorithm following Eqs.~\eqref{dis1}-\eqref{dis3} runs until the last time step $n=N$. It is suggested that the grid size $d\tau$ be chosen carefully. For an arbitrarily small $d\tau$, it may happen that $V_{i}=V_{i+1}$ which results in error. In this study, we use $d\tau$ which equals to the $\tau$ difference between the first and second frozen cells in the MSFMM algorithm. Next, we shall demonstrate and validate that the MSFMMC algorithm and DTOF-coordinate discretisation scheme is very efficient and accurate for simulating transient pressure at the wellbore. The simulation is conducted on the DTOF coordinate, and simulation results need to be mapped back to original 2D Cartesian grids only for visualisation. The entire workflow of using MSFMMC to compute DTOF and then using 1D DTOF-coordinate to solve the pressure equation can be named \textbf{MSFMMC-DTOF} for clarity.

\section{Validation Examples}
\subsection{A Homogeneous Model}
The first test case is a homogeneous oil reservoir model. There is a producing well in the centre of the reservoir by depletion. Suppose the oil reservoir is undersaturated with immobile water, then the production process can be modelled as slightly compressible single-phase flow. This assumption also applies to subsequent heterogeneous examples. The Peaceman's well model \citep{peaceman1978} is implemented to relate well-block pressure with bottom-hole pressure (BHP). Parameters for the reservoir and well model are summarised in Table~\ref{parameter}. The discretisation scheme using MSFMMC and 1D DTOF-coordinate is compared to standard FMM and finite difference (FD) methods. For FD, harmonic average of transmissibility, i.e. two-point flux approximation \citep{lie2014}, is used between grid cells.

\begin{figure}[!htp] 
	\centering
	\begin{subfigure}[t]{0.544\textwidth}
		\centering
		\includegraphics[width=1\textwidth]{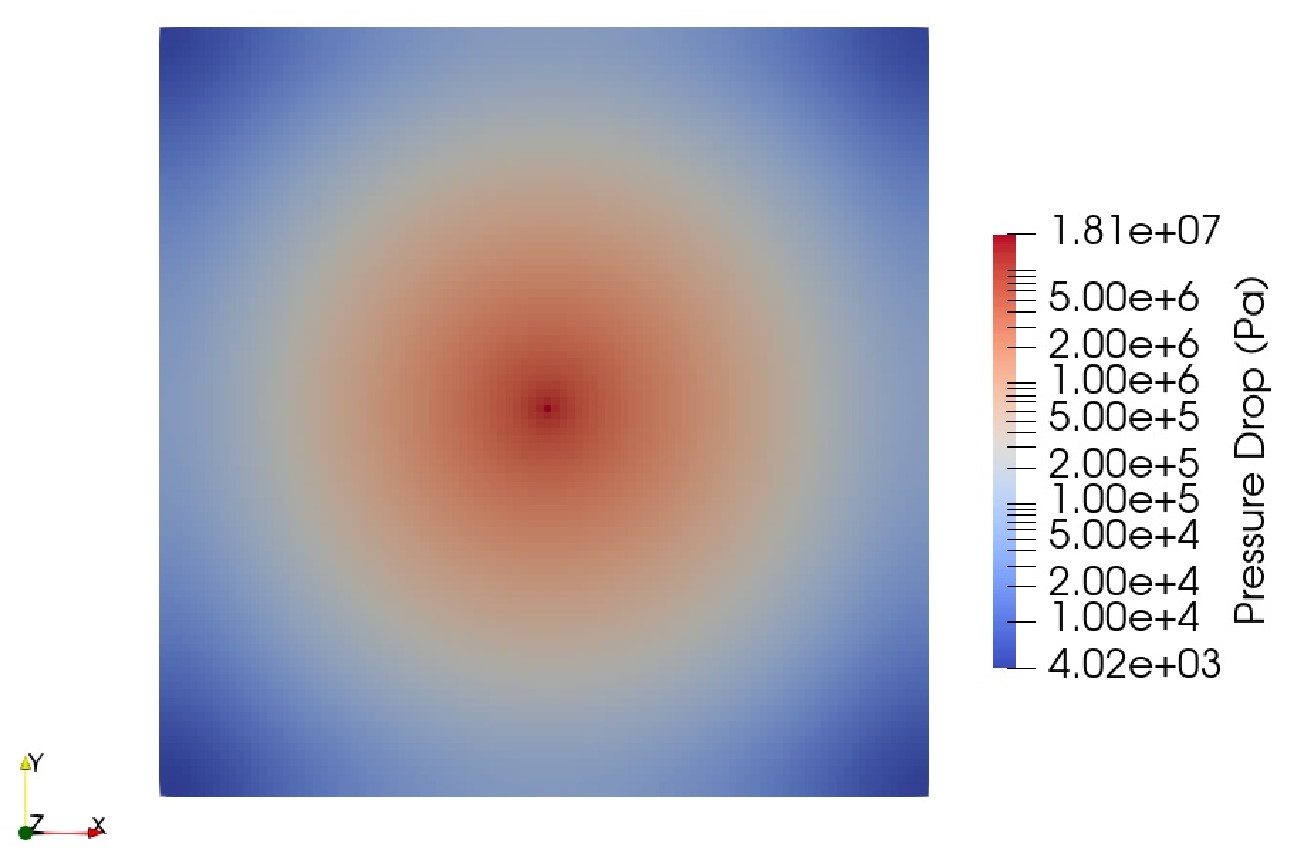}
		\caption{}
		\label{presshomo} 
	\end{subfigure}
	\begin{subfigure}[t]{0.44\textwidth}
		\centering
		\includegraphics[width=1\textwidth]{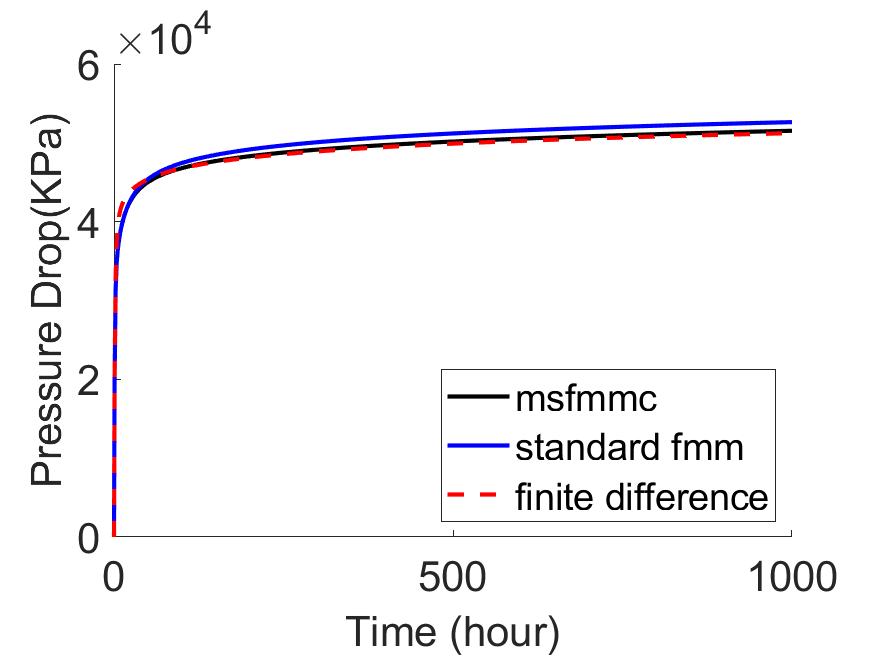}
		\caption{}
		\label{dpchomo}
	\end{subfigure}
	\caption{The pressure field at end of simulation for the homogeneous model (a); comparison of BHP drop for MSFMMC, FMM and FD on the homogeneous model (b).}
\end{figure}

The initial pressure of the reservoir is 50 mPa everywhere. The simulation is run for 1000 hours with a constant flow rate 86.4 m$^3$/day for the producer. The Cartesian grid for the finite difference method is 100$\times$100, while there are only 133 cells on the 1D DTOF-coordinate. The time step is 10s for both grids. Therefore, the computational efficiency is considerably increased by simulating on the DTOF-coordinate. Fig.~\ref{presshomo} shows the pressure field at the end of simulation for FD, and that for MSFMMC is almost identical. Fig.~\ref{dpchomo} compares the change of BHP drop with time for MSFMMC, FMM and FD. The BHP drop curves for MSFMMC and FD almost overlap while that for FMM is slightly higher. This test case validates the accuracy of MSFMMC and the DTOF-coordinate discretisation scheme on a homogeneous case. 
\begin{table}[!htp]
	\centering
	\begin{tabular}{|l|l|}
		\hline
		Parameter & Value \\
		\hline
		Porosity & 10\%\\
		\hline
		Effective Permeability for Oil & 5 mD\\
		\hline
		Total Compressibility & $8\times 10^{-9}$ Pa$^{-1}$\\
		\hline
		Dynamic Viscosity & 0.001 Pa*s\\
		\hline
		Initial Pressure & 50 mPa\\
		\hline
		Wellbore Radius & 0.05 m\\
		\hline
		Skin factor & 5\\
		\hline
		Wellbore storage coefficient & $1.256\times 10^{-7}$ m$^3$Pa$^{-1}$\\
		\hline
		Formation Volume Factor & 1.2\\
		\hline
		Well Producing Rate & 86.4 m$^3$/day\\
		\hline
	\end{tabular}
	\caption{A summary of reservoir and well model parameters.}
	\label{parameter}
\end{table}

\subsection{A Heterogeneous Model-Single Channel}
\begin{figure}[!htp] 
	\centering
	\begin{subfigure}[t]{0.4\textwidth}
		\centering
		\includegraphics[width=1\textwidth]{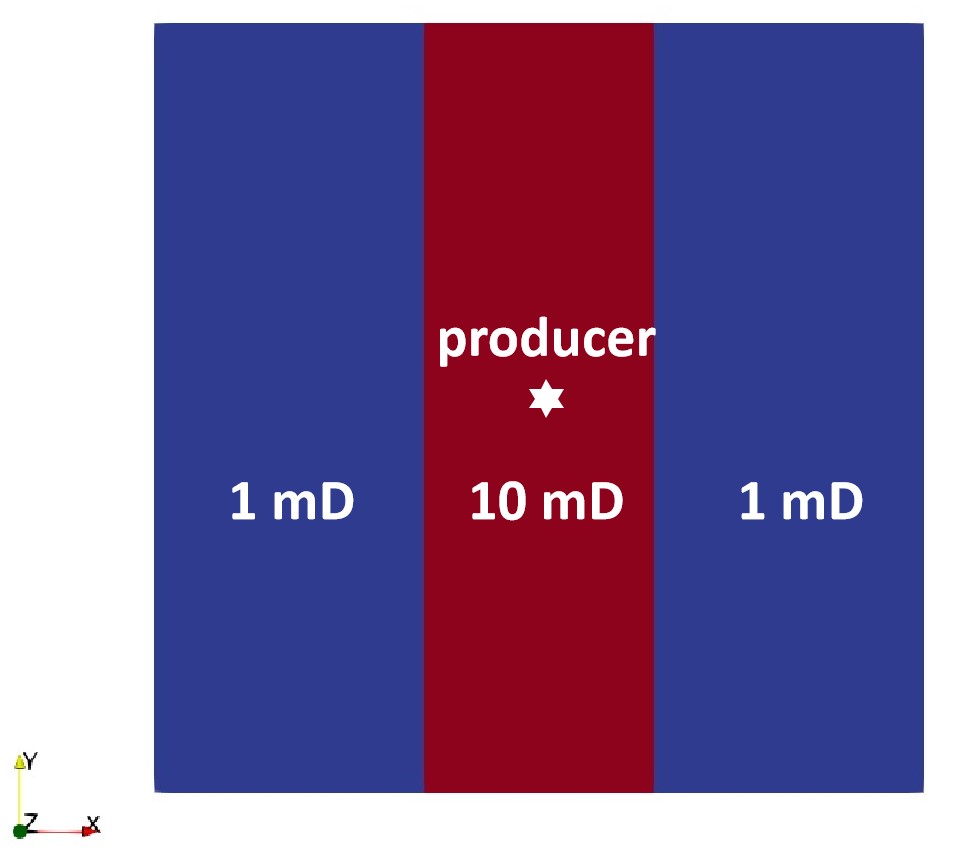}
		\caption{}
		\label{heteperm} 
	\end{subfigure}
	\begin{subfigure}[t]{0.544\textwidth}
		\centering
		\includegraphics[width=1\textwidth]{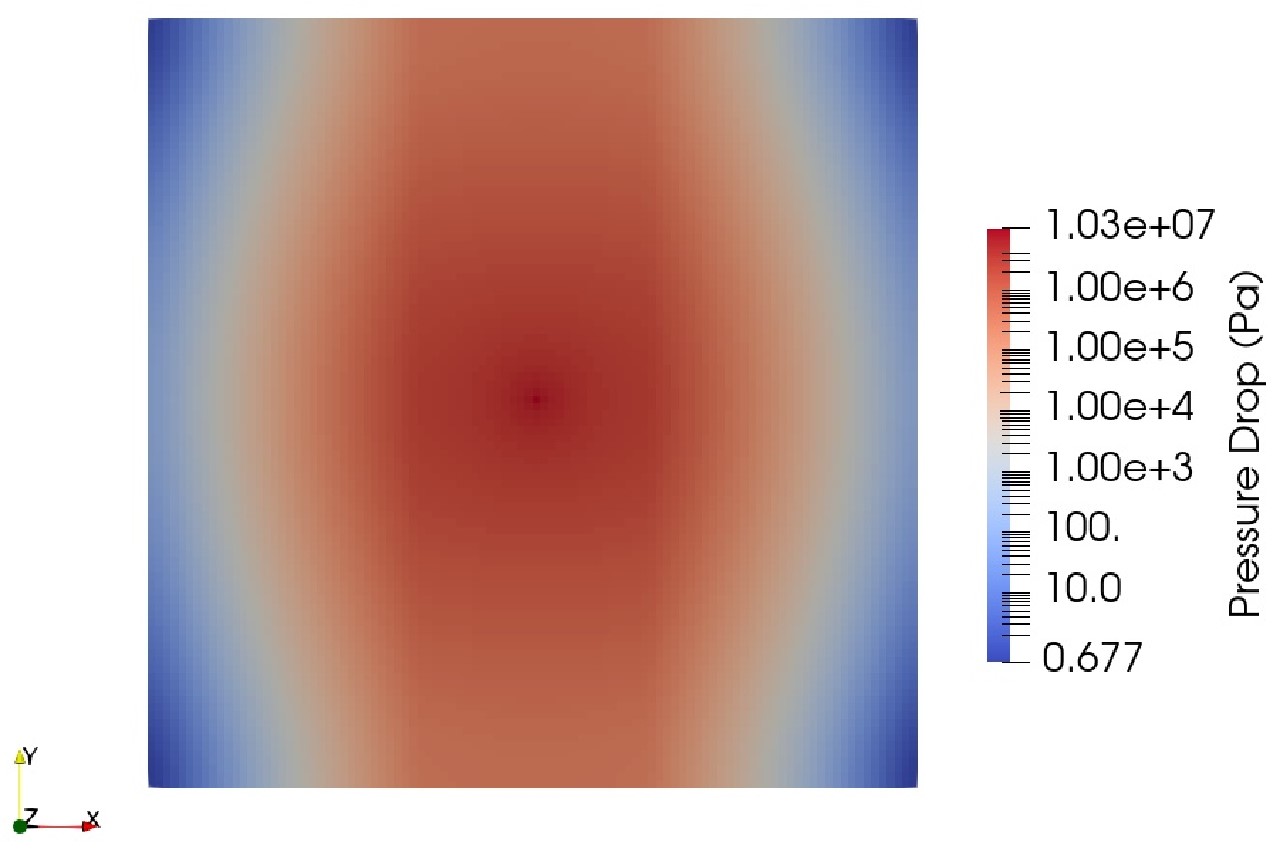}
		\caption{}
		\label{presshete}
	\end{subfigure}
	\caption{The effective permeability field for the heterogeneous single channel model (a); the corresponding pressure drop field at end of simulation (b).}
\end{figure}

\begin{figure}[!htp]
	\centering
	\includegraphics[width=0.6\linewidth]{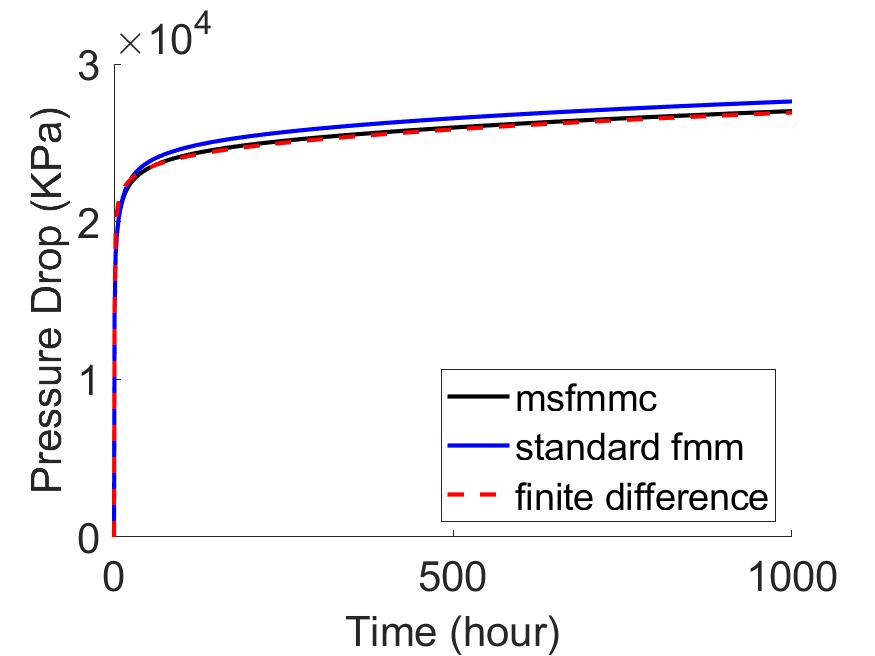}
	\caption{Comparison of BHP drop for MSFMMC, FMM and FD on the single channel model.}
	\label{dpchete}
\end{figure}

Next, a heterogeneous model is used to validate our algorithm. The effective permeability field is in Fig.~\ref{heteperm} and other parameters are the same as in Table~\ref{parameter}. The Cartesian grid for the finite difference method is 100$\times$100, while there are only 164 cells on the 1D DTOF-coordinate. The time step is 10s for both grids. The simulated pressure fields at the end of 1000 hour of MSFMMC and FD are very close and shown in Fig.~\ref{presshete}. The influence of the high-permeability channel on pressure is clearly seen. The change of BHP drop with time for MSFMMC, FMM and FD are compared in Fig.~\ref{dpchete}. The pressure drop curves for FD and MSFMMC almost overlap with each other except in the initial 20 hours of producing.

\subsection{A Heterogeneous Model-Multiple Channels}
\begin{figure}[!htp] 
	\centering
	\begin{subfigure}[t]{0.4\textwidth}
		\centering
		\includegraphics[width=1\textwidth]{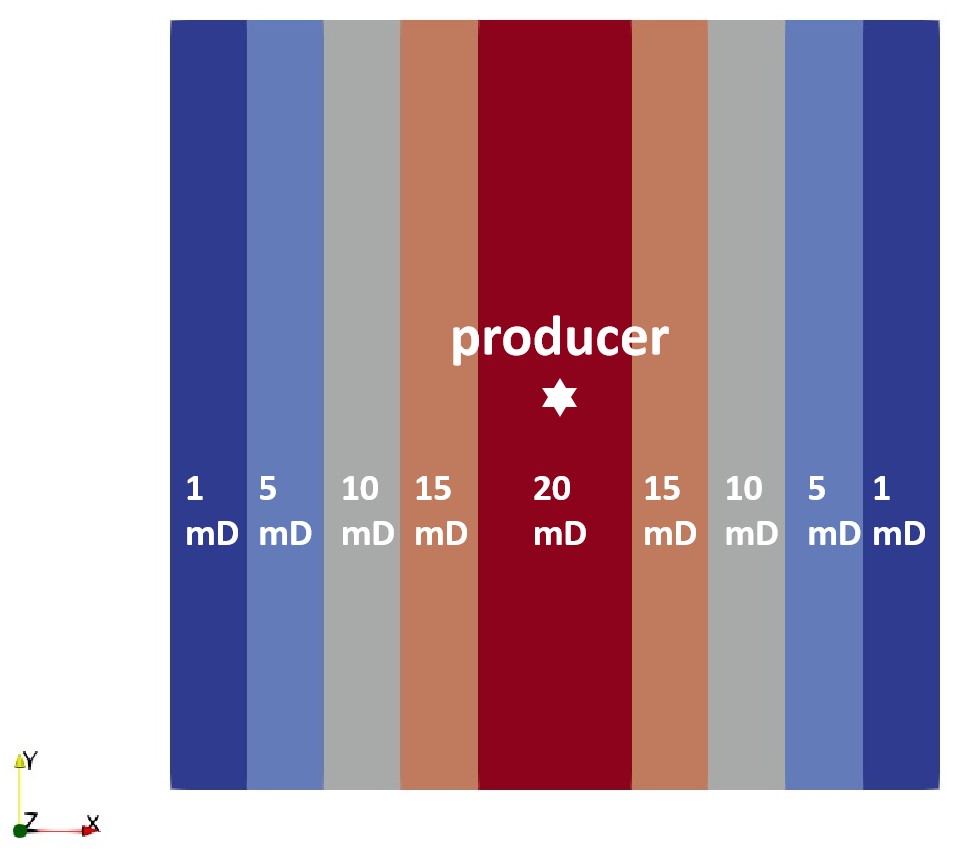}
		\caption{}
		\label{moreheteperm} 
	\end{subfigure}
	\begin{subfigure}[t]{0.544\textwidth}
		\centering
		\includegraphics[width=1\textwidth]{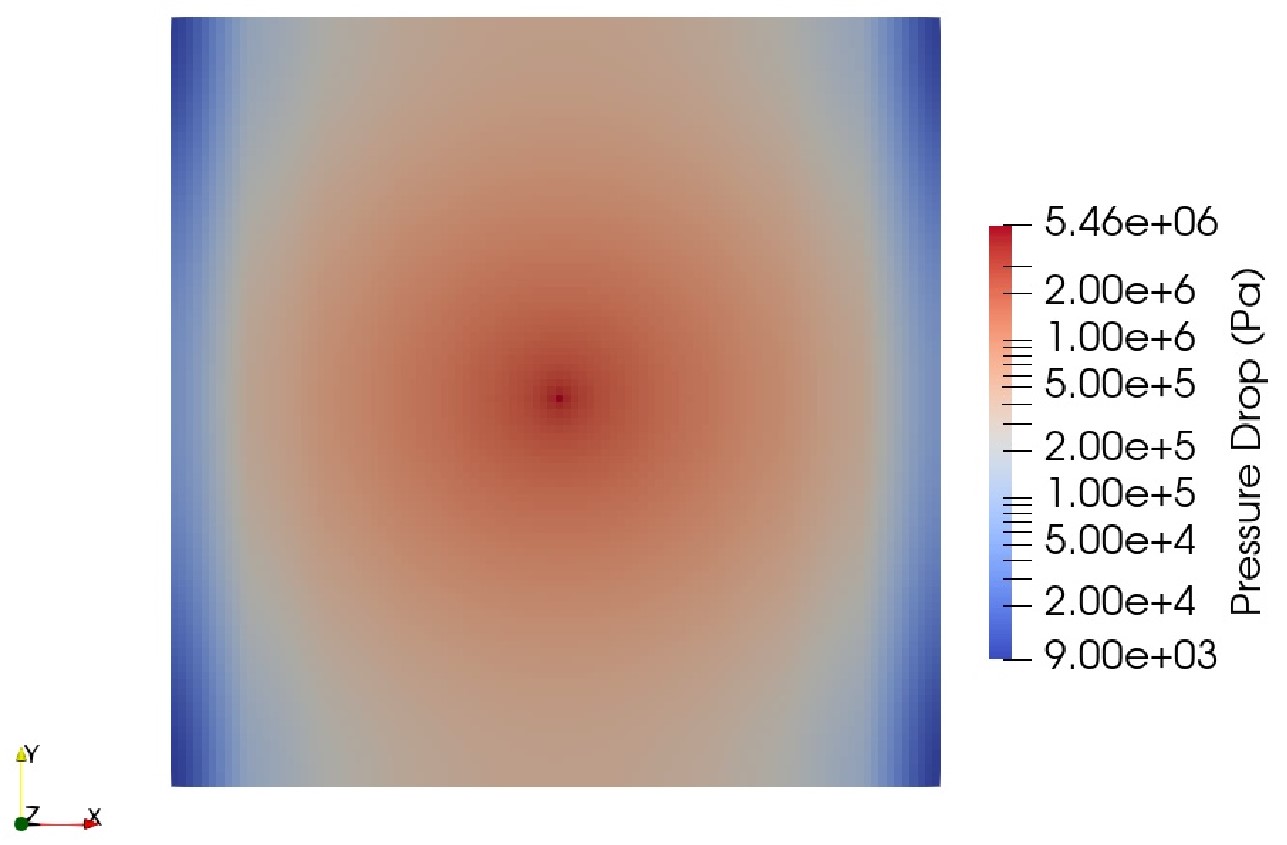}
		\caption{}
		\label{pressmorehete}
	\end{subfigure}
	\caption{The effective permeability field for the multiple-channels model (a); the corresponding pressure drop field at end of simulation (b).}
\end{figure}

\begin{figure}[!htp]
	\centering
	\includegraphics[width=0.6\linewidth]{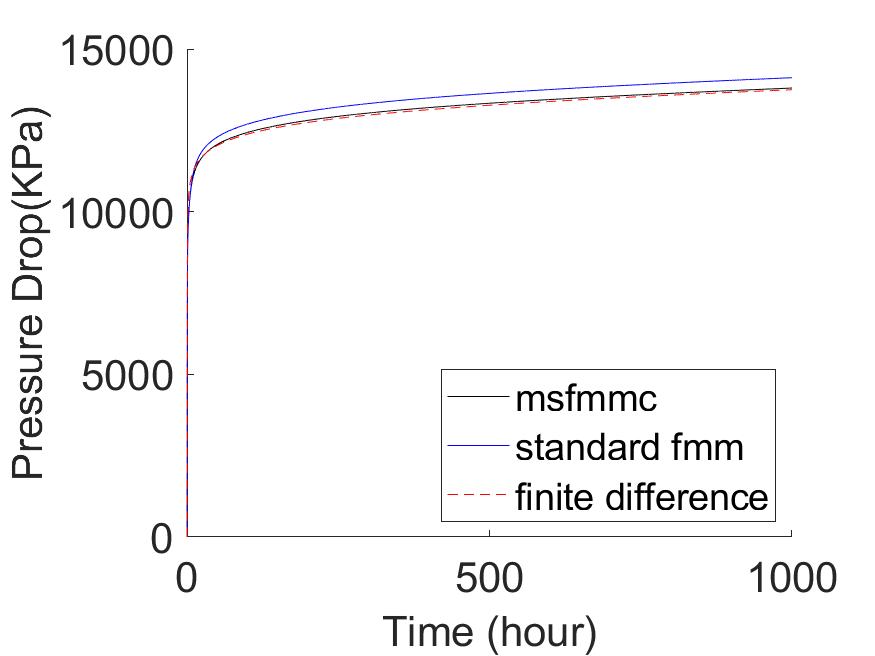}
	\caption{Comparison of BHP drop for MSFMMC, FMM and FD on the multiple-channels model.}
	\label{dpcmorehete}
\end{figure}

Next, we increase the heterogeneity by building more channels of different permeabilities shown in Fig.~\ref{moreheteperm}. All other reservoir and well parameters are the same as in Table~\ref{parameter}. The Cartesian grid for the finite difference method is 100$\times$100, while there are only 133 cells on the 1D DTOF-coordinate. The time step is 10s for both grids. The simulated pressure fields at the end of 1000 hour of MSFMMC and FD are very close and shown in Fig.~\ref{pressmorehete}. The change of BHP drop with time for MSFMMC, FMM and FD are compared in Fig.~\ref{dpcmorehete}. The pressure drop curves for FD and MSFMMC almost overlap with each other except in the initial 20 hours of producing.

\subsection{A Highly Heterogeneous Model}
\begin{figure}[!htp]
	\centering
	\includegraphics[width=0.6\linewidth]{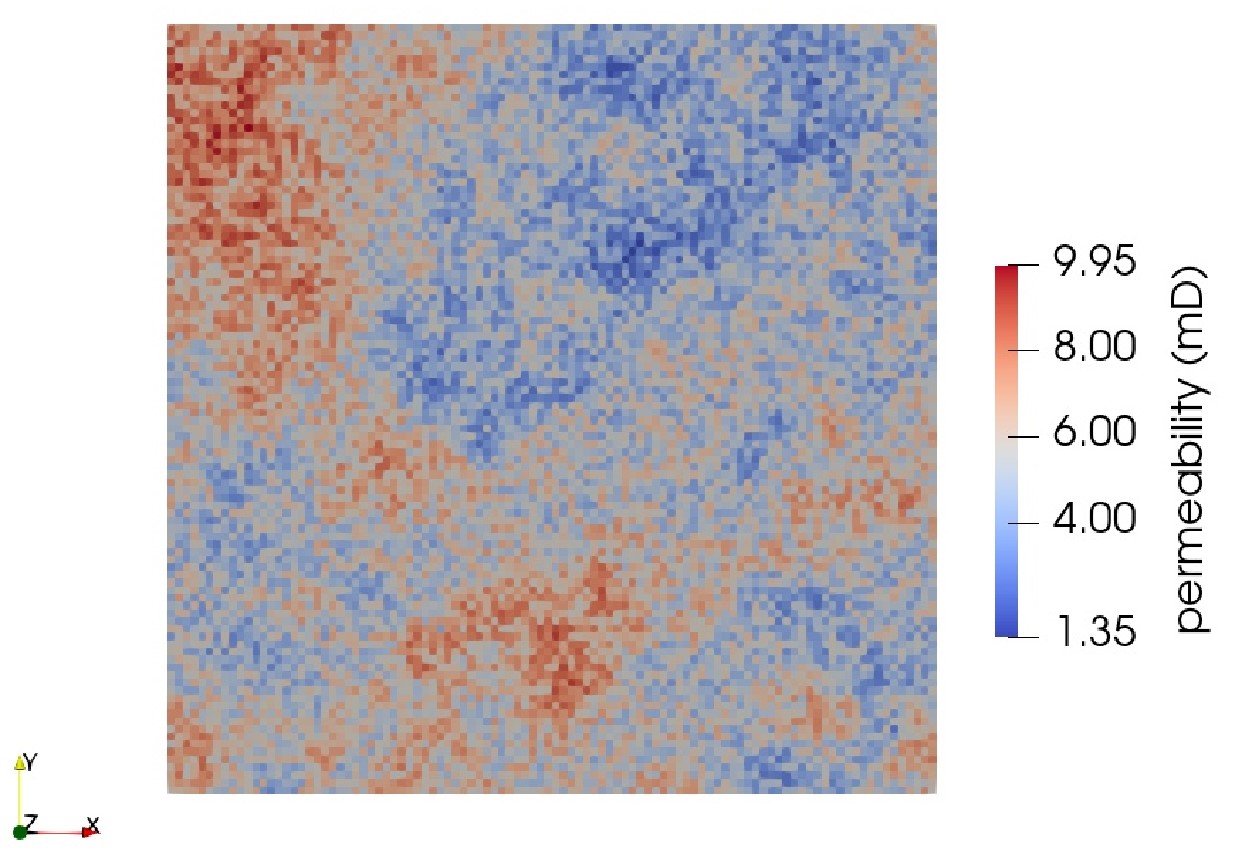}
	\caption{The effective permeability field for the highly heterogeneous model generated by sequential Gaussian simulation using SGeMS.}
	\label{targetperm}
\end{figure}

\begin{figure}[!htp] 
	\centering
	\begin{subfigure}[t]{0.48\textwidth}
		\centering
		\includegraphics[width=1\textwidth]{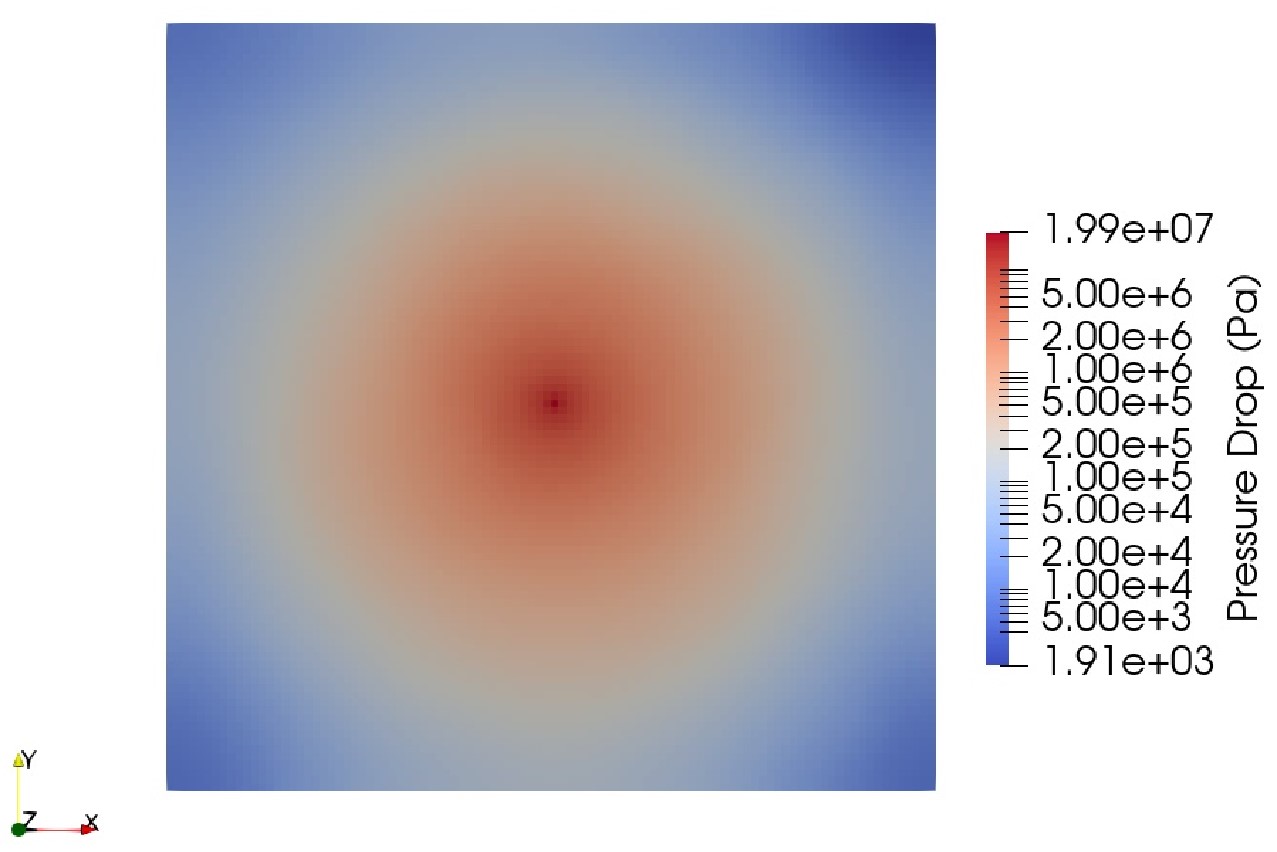}
		\caption{}
		\label{presstargetfd} 
	\end{subfigure}
	\begin{subfigure}[t]{0.48\textwidth}
		\centering
		\includegraphics[width=1\textwidth]{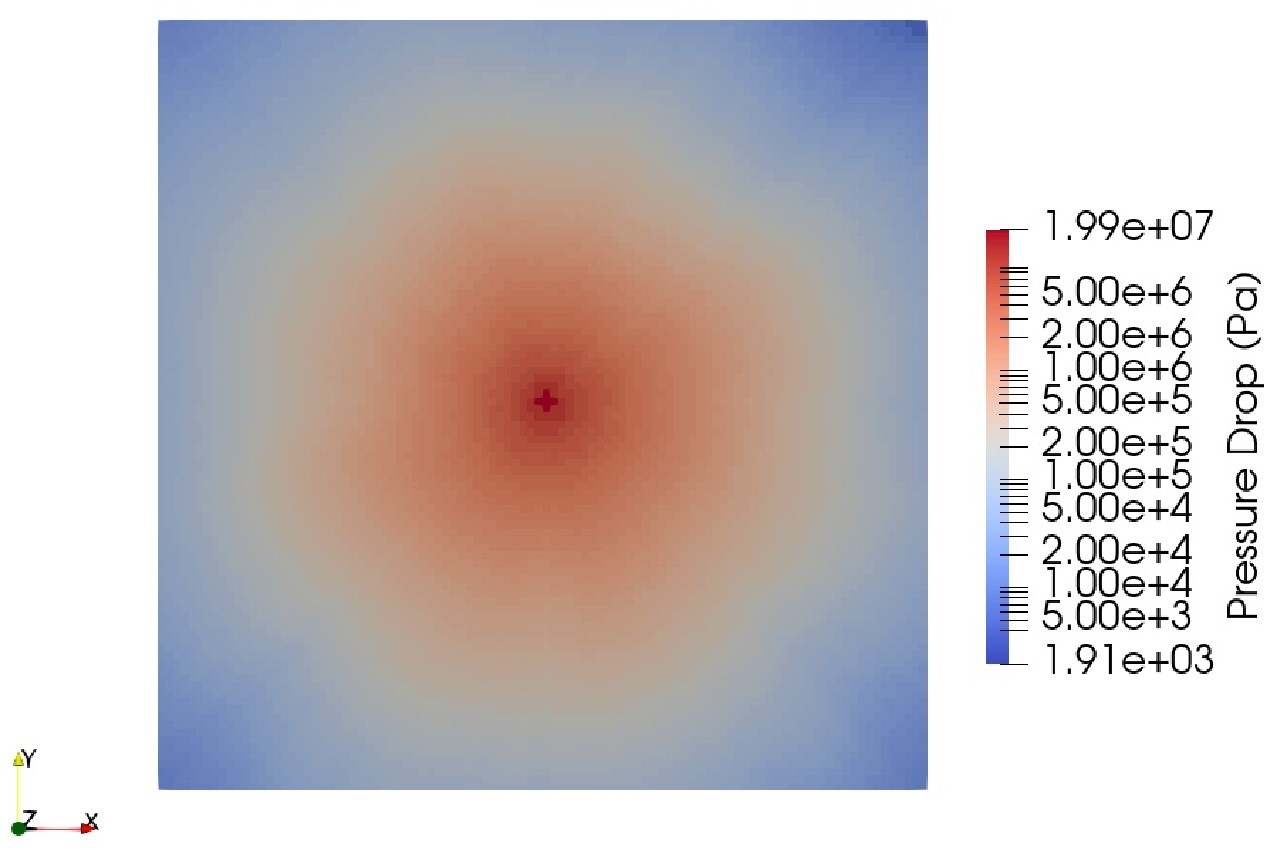}
		\caption{}
		\label{presstargetmsfmmc}
	\end{subfigure}
	\caption{The pressure drop field at end of simulation for the highly heterogeneous model using FD (a) and MSFMMC (b).}
\end{figure}

\begin{figure}[!htp]
	\centering
	\includegraphics[width=0.6\linewidth]{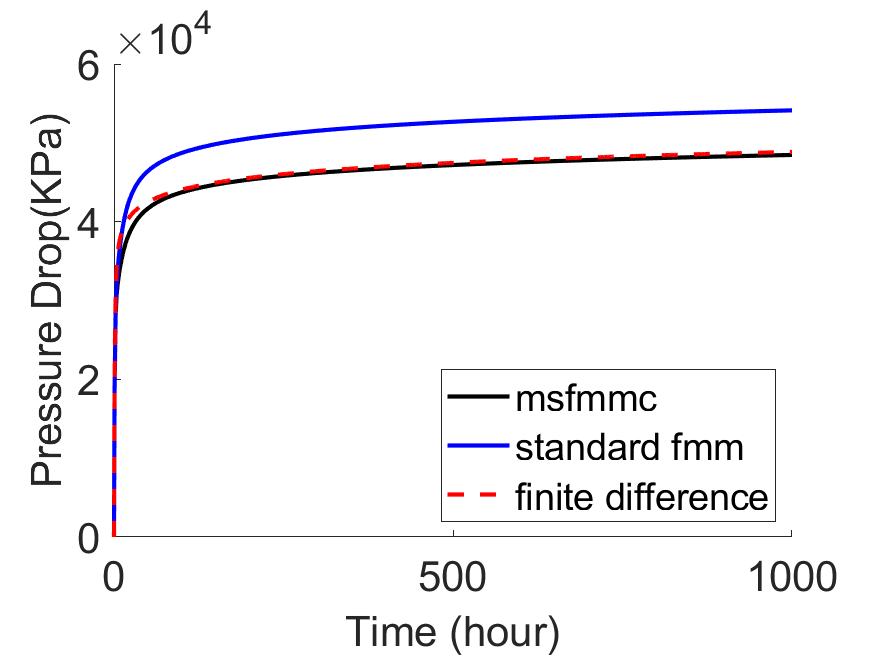}
	\caption{Comparison of BHP drop for MSFMMC, FMM and FD on the highly heterogeneous model.}
	\label{dpc_target}
\end{figure}

In this example, a highly heterogeneous reservoir model shown in Fig.~\ref{targetperm} is built by sequential Gaussian simulation using SGeMS on the Ely1 2D data set \citep{remy2009applied}. The original data is about formation thickness but here is interpreted as effective permeability for oil with unit mD. A producing well is placed in the centre of the reservoir. Apart from permeability, all other parameters are the same as in Table~\ref{parameter}. Figs.~\ref{presstargetfd} and \ref{presstargetmsfmmc} presents the pressure fields at the end of 1000h simulation by FD and MSFMMC, respectively. The two pressure fields are similar generally except near the boundaries of the reservoir. It is not surprising that the pressure fields by FD and MSFMMC are not exactly the same since simulation is conducted on the 1D DTOF-coordinate in MSFMMC and only mapped to the 2D Cartesian grid for visualisation. The purpose of using MSFMMC instead of FD is to simulate wellbore pressure efficiently. It can be seen in Fig.~\ref{dpc_target} that the BHP drops by MSFMMC and FD are almost the same except the initial 30 hours of producing. This validates our algorithm in a highly heterogeneous example.

\section{Potential Application in Automated History Matching}
In this section, we demonstrate the potential of MSFMMC in automated history matching (AHM). In fact, adopting MSFMMC in AHM is no different than using FD except that the computational efficiency can be considerably increased, which is a major advantage for AHM. The aim of AHM is to update the reservoir model such that the simulated dynamic responses can be as close to the observed data as possible. The objective function is defined to be
 
\begin{equation}
	M=\sum_{i=0}^{N}\frac{(obs_i-sim_i)^2}{2\sigma^2}~,\label{object}
\end{equation}

where $\sigma^2$ denotes the variance in the observed data. It can be proved that the likelihood $P(O|m)$ satisfies
\begin{equation}
	P(O|m)\propto e^{-M}~,
\end{equation}
and minimizing $M$ produces the maximum a posteriori estimate (MAP) for model parameters \citep{arnold2008geological}.
There have been many studies on optimization methods aiming at minimizing the object function. These methods can be grouped into gradient, stochastic and data assimilation methods \citep{valjak2008history, chen2020global, ma2020multiscale, ma2021Data,zhang2021history,yao2021optimization}. Stochastic methods are global optimization algorithms that prevents the entrapment in local minima and there is no need for computing gradients. These methods are more robust but the convergence rates can be slow. 
In the current study, the Differential Evolutionary (DE) optimization method \citep{hajizadeh2011population} is adopted to minimise the misfit. The steps of DE algorithm is
\begin{enumerate}
	\item Define and simulate the initial set of realisations, and compute the corresponding values of the objective function, i.e. the misfit. Rank the realisations in the set according to misfit. 
	\item Randomly select two realisations and calculate the difference vector which is subsequently added to a randomly selected base realisation to build a new realisation.
	\item Simulate the new realisation and compute the misfit.
	\item Add the new realisation to the set of realisations.
	\item Remove the realisation with highest misfit, rank the new set of realisations and calculate the new probability of selection for each realisation.
	\item Return to step 2 if the stopping criteria is not met. The stopping criteria is the misfit of all realisations in the set converge or a certain number of iterations is reached.
\end{enumerate}

\subsection{Simple Channel Model-Dimension 2}
\label{simple}
The first test case is the simple channel model in Fig.~\ref{heteperm} which is assumed to be real, and its BHP is assumed to be the measured data with observation interval 20 hours. The standard deviation $\sigma$ is assumed to be 10 KPa. Suppose the prior range for the permeability of the central channel (P1) is [2 mD, 20 mD], and that of the rest of the reservoir (P2) is [0.2 mD, 4 mD]. Other parameters are assumed to be known and kept unchanged. For P1, we sample 5 initial values (2 mD, 9 mD, 12 mD, 17 mD, 20 mD); for P2, we also sample 5 initial values (0.2 mD, 1.2 mD, 2.2 mD, 3.2 mD, 4 mD). P1 and P2 have 25 combinations in total, and the prior set of realisations are defined to have 25 members accounting for all combinations, i.e. the population size for DE is 25. This approach for defining the prior set of realisations applies for low-dimensional cases, i.e. when parameters for AHM are few. In this example, only two parameters P1 and P2 needs to be history matched, hence the dimension is 2. 

\begin{figure}[!htp]
	\centering
	\includegraphics[width=0.6\linewidth]{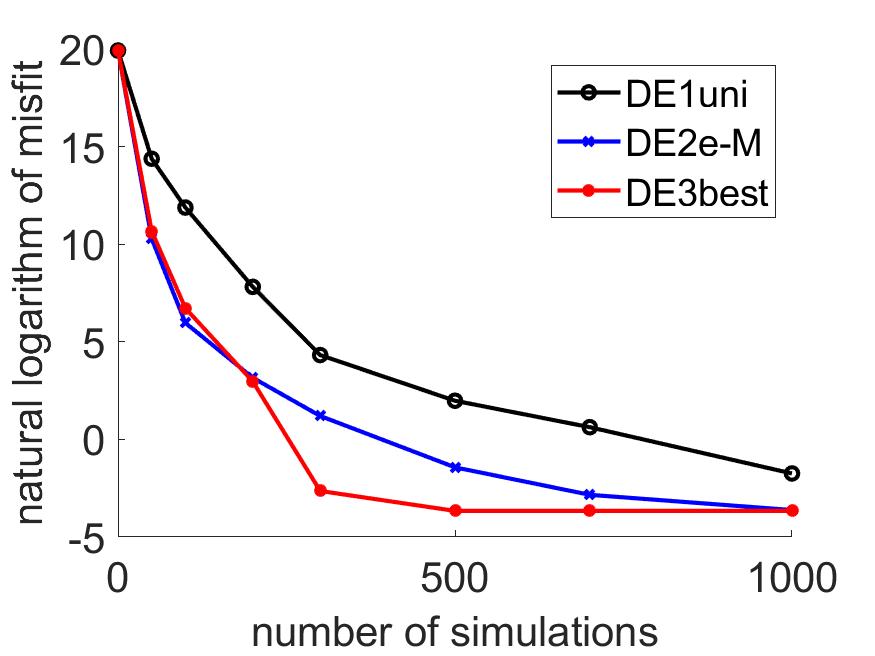}
	\caption{Convergence of misfit using different evolutional strategies for the AHM example of dimension 2.}
	\label{misfit_simple}
\end{figure}

In the process of AHM, the members in the set are replaced by new realisations while the total number of realisations in the set is kept constant. The number of simulations for updating realisations is set to be 1000. The strategy for evolution is an important factor that affects the performance of DE algorithm \citep{hajizadeh2011population}. By strategy, we mean Step 2 in the DE algorithm that can be written as

\begin{equation}
	r_{new}=r_{base}+s(r_1-r_2)~,
\end{equation}
where $r_{new}$ is the new realisation, $r_{base}$, $r_1$ and $r_2$ are three randomly selected realisations in the set.
In this example, we compare three different strategies, DE1uni, DE2e-M and DE3best. For DE1uni, $r_{base}$, $r_1$ and $r_2$ are all randomly selected assuming realisations in the set follow uniform distribution. For DE2e-M,  $r_{base}$ is chosen assuming the probability of selecting realisations is $\propto e^{-M}$. For DE3best, $r_{base}$ is chosen to be the realisation with the lowest misfit directly. Fig.~\ref{misfit_simple} shows the decrease of misfits with simulations for the three strategies. DE3best has the highest convergence rate. It is interesting to note that initially DE2e-M converges faster than DE3best. A possible explanation is that since the new realisation is built randomly, DE2e-M happens to generate realisations with lower misfits than DE3best in the initial simulations, although generally the convergence rate by DE3best is the highest. The MAP estimates for P1 and P2 are 1.018 mD and 9.999 mD, respectively, which are very accurate given the 'true' values are 1 mD and 10 mD for this validation example. The time cost for 1000 simulations is less than 10 min on a normal desktop PC which demonstrates the advantage of adopting MSFMMC in AHM.
 
\subsection{Multiple Channels Model-Dimension 6}
\begin{figure}[!htp]
	\centering
	\includegraphics[width=0.6\linewidth]{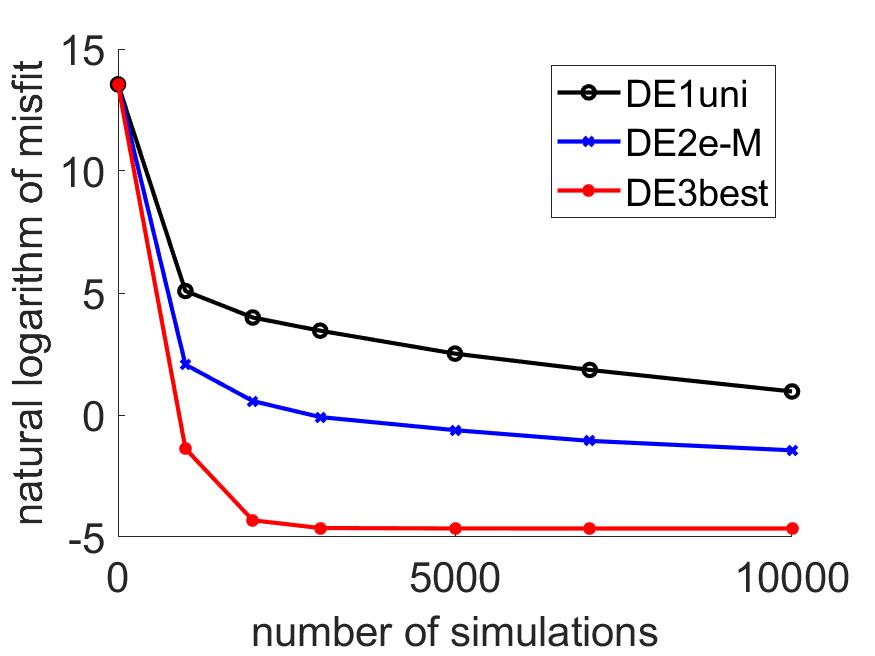}
	\caption{Convergence of misfit using different evolutional strategies for the AHM example of dimension 6. DE3best converged after 3000 simulations.}
	\label{misfit_complex}
\end{figure}

\begin{figure}[!htp]
	\centering
	\includegraphics[width=0.6\linewidth]{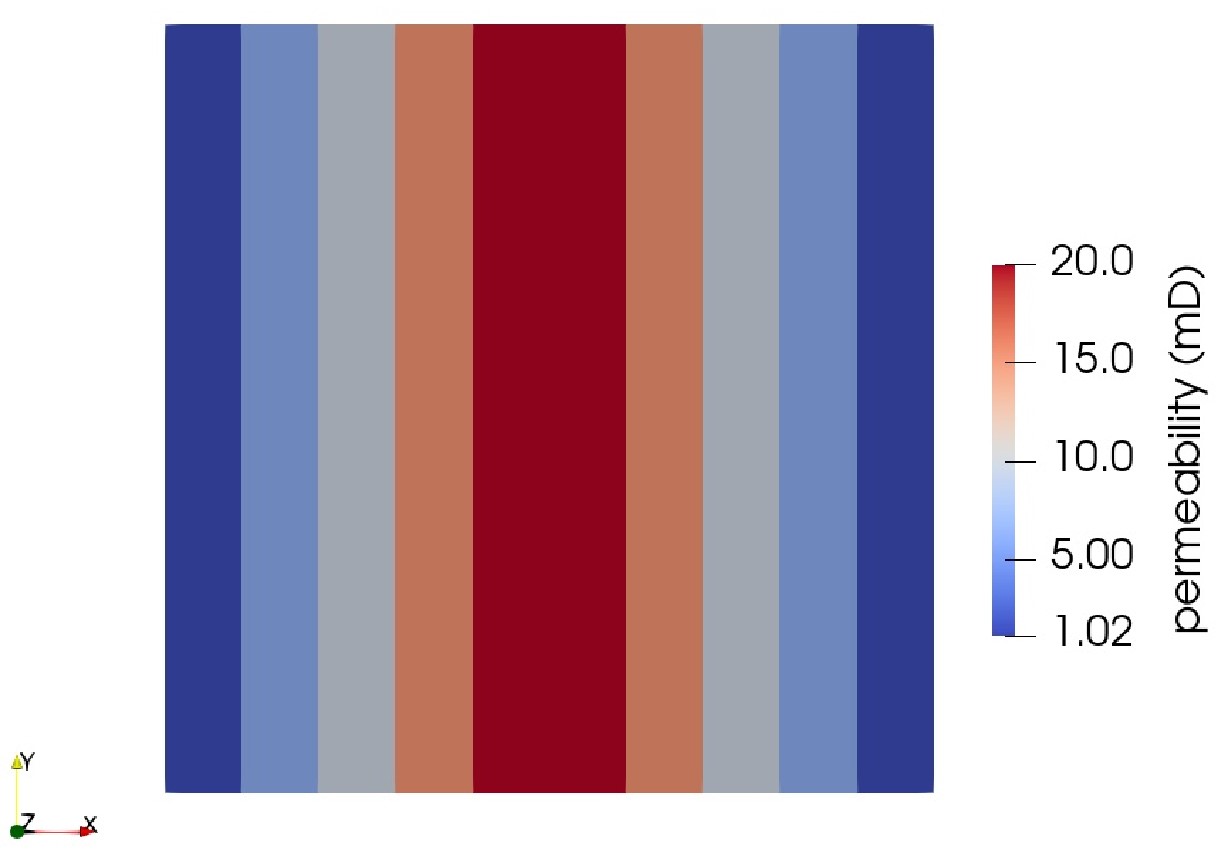}
	\caption{The MAP estimate for the permeability field after AHM.}
	\label{permahm}
\end{figure}

\begin{figure}[!htp]
	\centering
	\includegraphics[width=0.6\linewidth]{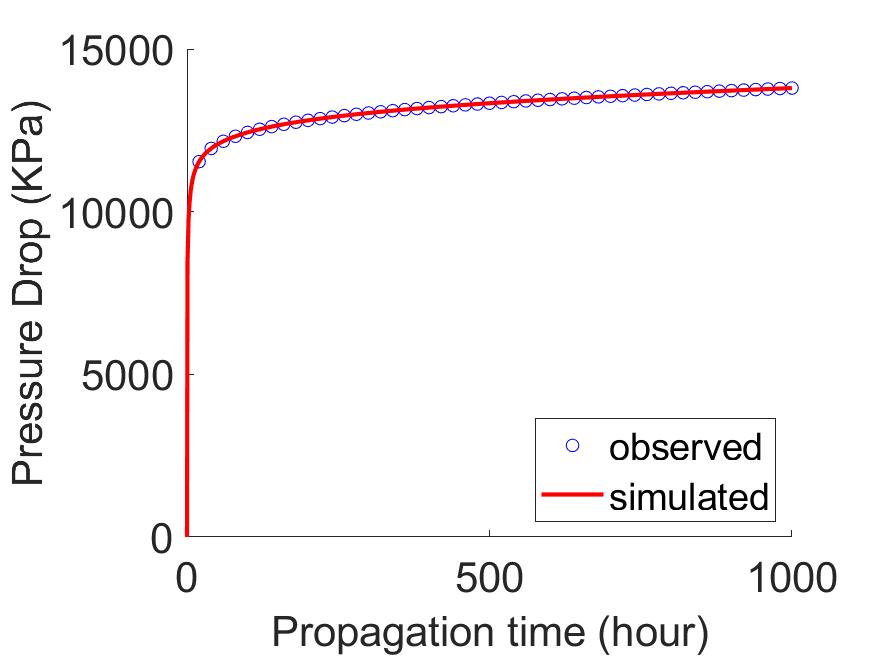}
	\caption{Comparison of observed and simulated BHP drop change with time after AHM.}
	\label{dpc_ahm_complex}
\end{figure}

Next, we demonstrate the potential of MSFMMC for AHM using an example of dimension 6. Suppose the 'realistic' reservoir model and parameters are the same as in Fig.~\ref{moreheteperm} consisting of channels of 5 different permeabilities and Table~\ref{parameter}. Its BHP is assumed to be the measured data with observation interval 20 hours. The permeabilities of channels are denoted as P1-P5 from the centre to boundary of the reservoir. In addition, well skin factor (P6) is assumed to be uncertain, hence, there are 6 parameters in total to be history matched. Their prior ranges are in Table~\ref{range}. Other parameters are the same as in Table~\ref{parameter}. The standard deviation $\sigma$ is assumed to be 10 KPa. The population size is set to be 100 and the prior realisations are generated by selecting parameters from its prior range randomly assuming uniform distribution. The same three evolutional strategies as in Section~\ref{simple} are compared in Fig.~\ref{misfit_complex}. For each strategy, a total of 10k simulations are conducted. It is obvious that DE3best has the highest convergence rate. The MAP estimate of skin factor and permeability field are 4.98 and Fig.~\ref{permahm}, respectively, which is close to the 'realistic' data. Corresponding to the low misfit after convergence, the simulated pressure drop is close to the 'measured' data (Fig.~\ref{dpc_ahm_complex}). For 10k simulations, the time cost is only about 1 hour, demonstrating the advantage of MSFMMC for forward simulation in AHM. 

It is worth mentioning that the perfect match of simulated and observed data, as well as the accurate MAP estimate in the current study, is partially due to the fact that synthetic 'realistic' data is used such that other parameters than the ones being history matched are known. For realistic problems, the dimension of the inversion problem may be much higher, the physical processes may not be well described by governing equations \citep{guo2019comprehensive}, and the parameters assumed to be known may not be accurate. These all make accurate AHM of practical cases not easy. In addition, the problem of multiplicity for high-dimensional inversion problems might become severe and regulations in the misfit calculation could be introduced  \citep{oliver2011recent}.

\begin{table}[!htp]
	\centering
	\begin{tabular}{|l|l|}
		\hline
		Parameter & Prior Range \\
		\hline
		permeability P1 & $[0.8, 1.2]$ mD\\
		\hline
		permeability P2 & $[4, 6]$ mD\\
		\hline
		permeability P3 & $[8, 12]$ mD\\
		\hline
		permeability P4 & $[12, 18]$ mD\\
		\hline
		permeability P5 & $[16, 24]$ mD\\
		\hline
		well skin factor P6 & $[4, 6]$\\
		\hline
	\end{tabular}
	\caption{Prior ranges of parameters for history matching.}
	\label{range}
\end{table}

\newpage
\section{Conclusions}
The efficiency of reservoir simulation is important for AHM and production optimisation. FMM-DTOF is an efficient method for reservoir simulation where the Eikonal equation of DTOF is solved by FMM and then the flow equation is computed efficiently on the 1D DTOF coordinate. In the current study, a new MSFMMC algorithm for solving DTOF and a discretisation algorithm for the flow equation on the 1D DTOF coordinate has been developed. The new algorithm has been validated on both homogeneous and heterogeneous examples showing that the BHP solution by MSFMMC-DTOF is more accurate than standard FMM-DTOF. In addition, the potential of MSFMMC-DTOF for efficient forward simulation in AHM has been demonstrated on two inversion problems of dimensions 2 and 6. MSFMMC-DTOF is a promising alternative for conventional FD or finite element-based reservoir simulators for many AHM problems involving slightly compressible single-phase flows, such as undersaturated oil reservoir production by depletion with immobile water. Further developments are needed for multiphase and compressible flows to allow more applications.

\section*{Acknowledgements}
The research is supported by Open Fund (PLN201918) of State Key Laboratory of Oil and Gas Reservoir Geology and Exploitation (Southwest Petroleum University) and National Science Fund for Distinguished Young Scholars (51525404).
\newpage
\section*{Bibliography}



  \bibliographystyle{elsarticle-harv} 
  \bibliography{mybib}



\newpage

\end{document}